\documentclass[fleqn,usenatbib,useAMS]{mnras}

\usepackage{graphicx}	
\usepackage{amsmath}	
\usepackage{amssymb}	
\usepackage{lmodern}    
\usepackage{multicol}     
\usepackage{bm}		     
\usepackage{pdflscape}	
\usepackage{xcolor}

\usepackage[T1]{fontenc}
\usepackage{ae,aecompl}
\usepackage{newtxtext,newtxmath}

\title[Say NIHAO to black holes]{NIHAO XXII: Introducing black hole formation, accretion and feedback into the NIHAO simulation suite}
\author[M. Blank, A.V. Macci\`o, A. A. Dutton, A. Obreja]{Marvin Blank$^{1,2}$\thanks{marvin.blank@nyu.edu}, Andrea V. Macci\`o$^{1,3}$, Aaron A. Dutton$^{1}$ and Aura Obreja$^{4,1}$ \\
$^{1}$New York University Abu Dhabi, PO Box 129188, Saadiyat Island, Abu Dhabi, United Arab Emirates\\
$^{2}$Institut f\"{u}r Theoretische Physik und Astrophysik, Christian-Albrechts-Universit\"{a}t zu Kiel, Leibnizstr. 15, D-24118 Kiel, Germany\\
$^{3}$Max Planck Institut f\"{u}r Astronomie, K\"{o}nigstuhl 17, D-69117 Heidelberg, Germany\\
$^{4}$Universit\"{a}ts-Sternwarte, Ludwig-Maximilians-Universit\"{a}t M\"{u}nchen, Scheinerstr. 1, D-81679 Munich, Germany}

\date{\today}

\pubyear{2019}

\begin{document}
\label{firstpage}
\pagerange{\pageref{firstpage}--\pageref{lastpage}}
\maketitle

\begin{abstract}
We introduce algorithms for black hole physics, i.e., black hole formation, accretion and feedback,
into the NIHAO (Numerical Investigation of a Hundred Astrophysical Objects) project of galaxy simulations.
This enables us to study high mass, elliptical galaxies, where feedback from the
central black hole is generally thought to have a significant effect on their evolution.
We furthermore extend the NIHAO suite by 45 simulations that encompass $z=0$ halo masses from $1 \times 10^{12}$ to  $4 \times 10^{13}\,\mathrm{M}_{\sun}$,
and resimulate five galaxies from the original NIHAO sample with black hole physics,
which have $z=0$ halo masses from $8 \times 10^{11}$ to $3 \times 10^{12}\,\mathrm{M}_{\sun}$.
Now NIHAO contains 144 different galaxies and thus has the largest sample of zoom-in simulations of galaxies,
spanning $z=0$ halo masses from $9 \times 10^{8}$ to $4 \times 10^{13}\,\mathrm{M}_{\sun}$.
In this paper we focus on testing the algorithms and calibrating their free parameters against the
stellar mass versus halo mass relation and the black hole mass versus stellar mass relation.
We also investigate the scatter of these relations, which we find is a decreasing function with time
and thus in agreement with observations.
For our fiducial choice of parameters we successfully quench star formation in objects above a $z=0$ halo mass of $10^{12}\,\mathrm{M}_{\sun}$,
thus transforming them into red and dead galaxies.
\end{abstract}



\section{Introduction}

It is now well established that black holes exist in the centres of almost all galaxies \citep[e.g.,][]{1995_Kormendy_Richstone,1998_Magorrian_Tremaine_Richstone}, and that they are connected to their hosts.
The general idea outlined by \citet{2005_DiMatteo_Springel_Hernquist} is that feedback provided by the central black hole heats the galaxy's gas and subsequently quenches star formation and further black hole accretion, transforming it into a \lq red and dead\rq\, elliptical galaxy.
In this framework this co-evolution leads the galaxy, i.e. its velocity dispersion \citep[e.g.,][]{2000_Ferrarese_Merritt,2000_Gebhardt_Bender_Bower} or its bulge mass \citep[e.g.,][]{1995_Kormendy_Richstone,2004_Haering_Rix}, being related to the black hole mass.
However, some works \citep[e.g.,][]{2011_Jahnke_Maccio} argue that correlations between the galaxy and its central black hole
do not imply a physical connection of these two.


Numerical simulations have been proven very successful in investigating the formation and evolution of galaxies, leading to a number of projects that have been developed over the last years.
E.g., Illustris \citep{2014_Vogelsberger_Genel_Springel}, IllustrisTNG \citep{2018_Pillepich_Springel_Nelson}, Magneticum Pathfinder \citep{2016_Dolag_Komatsu_Sunyaev}, MassiveBlack-II \citep{2015_Khandai_DiMatteo_Croft} and EAGLE \citep{2015_Schaye_Crain_Bower} are hydrodynamic simulations of cosmological volumes, and FIRE \citep{2014_Hopkins_Keres_Onorbe} and Auriga \citep{2016_Grand_Springel_Gomez} consist of zoom-in simulations of individual galaxies.


The NIHAO project \citep{2015_Wang_Dutton_Stinson} is a suite of hydrodynamical cosmological zoom-in simulations of galaxies and is unique in combining
(i) a high resolution of $\sim 10^6$ particles per halo,
(ii) a large range of halo masses from dwarf to Milky Way masses ($\sim 5 \times 10^{9}$ to $\sim 2 \times 10^{12} \, \rmn{M}_{\sun}$), and
(iii) a large sample size of $\sim$ 100 galaxies.
NIHAO successfully reproduces the stellar mass versus halo mass relation, one of the most fundamental constraints in galaxy formation and evolution, from redshift $z=0$ up to redshift $z=4$, and also matches the star formation rate versus stellar mass relation.
However, up to this date the NIHAO suite did not include high mass, elliptical galaxies, because the stellar feedback implemented into NIHAO is insufficient to regulate star formation in elliptical galaxies \citep{2015_Dutton_Maccio_Stinson}.
It is now generally accepted \citep{2006_Croton_Springel_White} that black hole feedback is needed at higher masses to reproduce the sharp decline in the stellar mass function and to create red and dead galaxies.


The aim of this paper is to introduce algorithms for black hole formation, accretion and feedback into the NIHAO project, and to extend the NIHAO suite to include high mass, elliptical galaxies.
We focus on the testing of the algorithms and the calibration of the free parameters on
the observed stellar mass versus halo mass and black hole mass versus stellar mass relations, and also investigate how the scatter of these relations evolve with time and compare with observations.
In section \ref{sec:nihao} we review the properties of the NIHAO project.
We introduce the algorithms used for black hole formation, accretion and feedback in section \ref{sec:numerics} and the initial conditions in section \ref{sec:ics}.
In section \ref{sec:results} we present our results and in section \ref{sec:paramstudy} we study the effect of the algorithms' free parameters on the simulations.
In section \ref{sec:summary} we summarize our findings.

\section{The NIHAO project}\label{sec:nihao}

NIHAO uses an updated version \citep{2014_Keller_Wadsley_Benincasa} of the TreeSPH code {\sc Gasoline2} \citep{2004_Wadsley_Stadel_Quinn, 2017_Wadsley_Keller_Quinn}.
The simulations use a flat LCDM cosmology with parameters from the \citet{2014_Planck_Collaboration}.

The backbone of the NIHAO project consists of cosmological dark-matter-only simulations with box sizes of 60 and 20 Mpc/h from \citet{2014_Dutton_Maccio}, and a new box of 15 Mpc/h with $400^3$ particles.
These are evolved until $z=0$, then haloes from these boxes are selected and resimulated individually with a higher resolution and with gas particles.

All galaxies have the same relative resolution across the whole mass range, i.e. $\sim 10^6$ dark matter particles inside the virial radius at $z=0$.
The particle masses and force softening lengths are chosen to resolve the mass profile at $\leq$ 1 per cent of the virial radius.
The initial ratio of dark and gas particle mass equals the cosmological mass ratio of dark matter and baryons of $\Omega_{\rmn{DM}}/\Omega_{\rmn{b}} = 5.48$.
The softening length of gas and star particles is $(\Omega_{\rmn{DM}}/\Omega_{\rmn{b}})^{1/2} = 2.34$ times smaller than the dark matter particle softening length.
The free parameters of the stellar and supernova feedback model have been chosen to match the $M_{\star}$-$M_{{200}}$ relation for one Milky Way-like galaxy at $z=0$.


Cooling is provided via hydrogen, helium, and various metal-lines in a uniform ultraviolet ionizing background \citep{2010_Shen_Wadsley_Stinson}, this includes photoionization, UV background heating \citep{2012_Haardt_Madau}, and Compton cooling.

Stars are formed from gas particles that pass a density and temperature threshold ($T < 15000 \, \rmn{K}$, $n > 10.3 \, \rmn{cm}^{-3}$) with a rate of $\dot{M_{\star}} = c_{\star} M_{\rmn{gas}} t_{\rmn{dyn}}^{-1}$, where $t_{\rmn{dyn}} = (4 \uppi G \rho)^{-1/2}$ is the gas particle's dynamical time, $\rho$ its density, $M_{\rmn{gas}}$ its mass and $c_{\star}=0.1$ the star formation efficiency.

Supernova feedback is modeled with the blastwave formalism of \citet{2006_Stinson_Seth_Katz}.
Stars of mass $8 < M_{\star}/M_{\sun} < 40$ eject metals and energy to surrounding gas particles 4 Myr after their formation.
For these gas particles cooling is delayed for $\sim 30 \, \rmn{Myr}$.
Before they produce a supernova massive stars provide \lq early stellar feedback\rq\, \citep{2013_Stinson_Brook_Maccio}, i.e. 13 per cent of the total stellar flux of $2 \times 10^{50} \, \rmn{erg} \, \rmn{M}_{\sun}^{-1}$ is injected into the surrounding gas as thermal energy.
No cooling delay is applied in this case.
We refer to \citet{2015_Wang_Dutton_Stinson} for more details of the NIHAO project.

NIHAO has been very successful in reproducing galaxy properties for halo masses of $M_{200} \leq 2 \times 10^{12}\,\rmn{M}_{\sun}$, e.g., the stellar mass versus halo mass relation \citep{2015_Wang_Dutton_Stinson}, the galaxy velocity function \citep{2016_Maccio_Udresco_Dutton}, the Tully-Fisher relation \citep{2017_Dutton_Obreja_Wang} and the rotation curves of dwarf galaxies \citep{2018_SantosSantos_DiCintio_Brook}.

\section{Computational methods for black hole physics}\label{sec:numerics}

Black holes are modeled as sink particles \citep{1995_Bate_Bonnell_Price} that only interact with their environment via gravitational forces and that can accrete matter from neighboring gas particles.
For black hole accretion and feedback we choose the models introduced by \citet{2005_Springel_Di-Matteo_Hernquist}, because these are the most widely used and thus tested, and are known to be able to yield the correct relation between black hole mass and the stellar component \citep{2005_DiMatteo_Springel_Hernquist}.

\subsection{Black hole formation}
We follow a common approach for modeling black hole formation: when a central halo\footnote{We do not seed subhaloes.}
exceeds a threshold mass $M_{\rmn{h,t}}$ we convert the gas particle (or a part thereof) with the lowest gravitational potential into a black hole with seed mass $M_{\rmn{BH,s}}$.\footnote{E.g., \citet{2007_Sijacki_Springel_DiMatteo} use $M_{\rmn{BH,s}} = 1 \times 10^5 \, \rmn{M}_{\sun}$ and $M_{\rmn{h,t}} = 5 \times 10^{10}$ M$_{\sun}$, \citet{2008_DiMatteo_Colberg_Springel} and \citet{2015_Schaye_Crain_Bower} use $M_{\rmn{BH,s}} = 1 \times 10^5 \, \rmn{M}_{\sun}$ and $M_{\rmn{h,t}} = 1 \times 10^{10}$ M$_{\sun}$, \citet{2010_Schaye_Dalla-Vecchia_Booth} use $M_{\rmn{BH,s}} = 9 \times 10^4 \, \rmn{M}_{\sun}$ and $M_{\rmn{h,t}} = 4 \times 10^{10} \, \rmn{M}_{\sun}$ and \citet{2015_Sijacki_Vogelsberger_Genel} use $M_{\rmn{BH,s}} = 1.4 \times 10^5 \, \rmn{M}_{\sun}$ and $M_{\rmn{h,t}} = 7.1 \times 10^{10} \, \rmn{M}_{\sun}$.}
Haloes and their masses are found with the AMIGA Halo Finder \citep[AHF,][]{2004_Gill_Knebe_Gibson,2009_Knollmann_Knebe}, see section \ref{sec:ics} for more details.
We follow \citet{2007_Sijacki_Springel_DiMatteo} and use the values $M_{\rmn{BH,s}} = 1 \times 10^5$ M$_{\sun}$ and $M_{\rmn{h,t}} = 5 \times 10^{10} \, \rmn{M}_{\sun}$.
An alternative model for black hole formation is used in the {\sc Romulus} simulations \citep{2017_Tremmel_Karcher_Governato}, where black hole formation occurs when the gas meets specific thresholds for metallicity, density and temperature.

\subsection{Black hole relocation}
In galaxy-scale simulations two effects can lead the position of the black hole not to coincide with the halo centre:
(i) The small ratio of black hole mass and gas/star/dark particle mass leads to stochastic motions of the black hole particle caused by the momentum of the accreted gas and gravitational interactions with nearby particles.
(ii) In the case of two haloes merging dynamical friction would lead the black hole to sink to the centre of the newly forming halo. However, in galaxy-scale simulations dynamical friction is generally underestimated on small scales due to insufficient resolution.

To compensate for these effects a number of models have been developed in the last years:
\citet{2011_Debuhr_Quataert_Ma,2012_Debuhr_Quataert_Ma} and \citet{2017_AnglesAlcazar_FaucherGiguere_Quataert} assign the black hole a \lq dynamical mass\rq\, or \lq tracer mass\rq, which is several orders of magnitude larger than the actual black hole mass, to prevent the black hole from moving too far away from the halo centre.
\citet{2009_Johansson_Naab_Burkert} relocate the black hole to the gas particle within the SPH smoothing length of the BH that has the lowest gravitational potential.
\citet{2009_Booth_Schaye} also relocate the black hole to the neighboring gas particle with the lowest gravitational potential, but only if the relative velocity of black hole and gravitationally most bound gas particle is smaller than 25 per cent of the local speed of sound and if the black hole mass is smaller than ten times the initial gas particle mass.
The IllustrisTNG simulations \citep{2015_Sijacki_Vogelsberger_Genel} force the black hole particle to be located at the potential minimum of its host halo.
The EAGLE simulations \citep{2015_Schaye_Crain_Bower} force black holes with a mass smaller than 100 times the gas particle mass to migrate towards the minimum of the gravitational potential of its host halo.
The Magneticum Pathfinder simulations \citep{2014_Hirschmann_Dolag_Saro} and the the {\sc Romulus} simulations \citep{2017_Tremmel_Karcher_Governato,2015_Tremmel_Governato_Volonteri} include a subgrid model to account for the dynamical friction force that is acting on the black hole.

However, relocating the black hole to the potential minimum of its host halo can lead it to move very large distances (the sum of the haloes virial radii, i.e., several $100\,\rmn{kpc}$) in the case of two haloes merging.
Furthermore, if these two haloes, whose black holes have just merged, disconnect again, the halo that just lost its black hole would get re-seeded.
To avoid these problems every major time step\footnote{The simulation time of 13.8 Gyr is divided into 1024 major time steps of 13.4 Myr.} we set the position and velocity of the black hole to the values of the dark matter particle within ten softening lengths that has the lowest gravitational potential.
Using the star particle with the lowest gravitational potential might be problematic at high redshifts, when only few or no star particles are present.
Placing the black hole at the gas particle with the lowest gravitational potential would artificially increase its kernel weight, thus its density and thus the black hole accretion rate. Furthermore, using a gas particle that has recently received feedback and is thus outflowing could lead to inappropriate black hole velocities and/or large jumps in the black hole position. \citep[See][for a discussion of a similar problem.]{2013_Wurster_Thacker}.

\subsection{Black hole merging}
Two black holes are merging when their distance is smaller than the sum of their softening lengths. The resulting black hole inherits the position, velocity and acceleration of its predecessors' barycentre, its mass is the sum of its predecessors' masses, and its accretion rate is calculated in the next timestep.

\subsection{Black hole accretion}
We calculate the accretion rate of the black hole with the commonly used\footnote{E.g., in \citet{2005_DiMatteo_Springel_Hernquist}, \citet{2005_Springel_Di-Matteo_Hernquist}, \citet{2008_Colberg_DiMatteo}, \citet{2008_DiMatteo_Colberg_Springel}, \citet{2009_Booth_Schaye}, \citet{2009_Croft_DiMatteo_Springel}, \citet{2009_Johansson_Naab_Burkert}, \citet{2012_Choi_Ostriker_Naab}.} Bondi-Hoyle-Lyttleton parametrization \citep{1939_Hoyle_Lyttleton, 1944_Bondi_Hoyle, 1952_Bondi}
\begin{equation}
  \dot{M}_{\rmn{BHL}} = \frac{4 \uppi \alpha G^2 M_{\rmn{BH}}^2 \rho}{(c_{\rmn{s}}^2 +v^2 )^{3/2}} \,,
  \label{eq:bondi}
\end{equation}
where $M_{\rmn{BH}}$ is the black hole mass and $\rho$, $c_{\rmn{s}}$ and $v$ are the density, sound speed and velocity of the gas that surrounds the black hole.
The parameter $\alpha$ was first introduced by \citet{2005_Springel_Di-Matteo_Hernquist} to account for the limited resolution of these simulations, and is usually set to $\alpha = 100$ \citep[see, e.g., table 2 in][]{2009_Booth_Schaye}.
However, our resolution is higher than in earlier works that use the Bondi-Hoyle-Lyttleton parametrization, therefore we use an accretion parameter of $\alpha = 70$ and justify this choice with a parameter study in section \ref{sec:paramstudy}.

The black hole accretion rate is limited by the Eddington rate \citep{1921_Eddington}
\begin{equation}
  \dot{M}_{\rmn{Edd}} = \frac{M_{\rmn{BH}}}{\epsilon_{\rmn{r}} \tau_{\rmn{S}}}
\end{equation}
with the Salpeter time-scale  $\tau_{\rmn{S}} = 4.5 \times 10^{8} \, \rmn{yr}$ \citep{1964_Salpeter} and the radiative efficiency $\epsilon_{\rmn{r}} = 0.1$ \citep{1973_Shakura_Sunyaev}.
The black hole accretion rate then is
\begin{equation}
  \dot{M}_{\rmn{BH}} = \min (\dot{M}_{\rmn{BHL}},\dot{M}_{\rmn{Edd}}) \,.
\end{equation}

At each time step $\Delta t$ the black hole accretes the mass $\Delta t \, \dot{M}_{\rmn{BH}}$ from the gas particle(s) that is (are) most gravitationally bound to the black hole. The momentum of the accreted mass is added to the black hole's momentum.
The black hole can accrete fractions of a gas particle, which might lead to gas particles with a very small mass.
If a gas particle's mass falls below 20 per cent of its initial mass it is deleted and its mass and momentum are distributed among the surrounding gas particles weighted with the SPH kernel.

To avoid too large accelerations and thus too small time steps of particles close to the black hole, as well as two body relaxation, we increase the black hole softening length as it grows.
As the softening length of collisionless particles in our simulations is proportional to the square root of their mass, we multiply the black hole softening length with $(1+ \Delta m/m_{\rmn{BH}})^{1/2}$ when it increases by $\Delta m$ in mass.

Increasing the softening length of the black hole does not significantly affect its surroundings, as its gravitational force is usually much smaller than the gravitational force of the stars surrounding the centre. E.g., for the galaxy g7.92e12 that we use for our parameter study in section \ref{sec:paramstudy}, the mass of the stars within $1\,\rmn{kpc}$ of the black hole is at least 25 times larger than the mass of the black hole for all times.

We do not limit how far particles can be away in order to be accreted. However, usually in each timestep not more than one particle is accreted, and this one is the most bound one, and thus usually the closest to the black hole.

Alternative methods to model gas accretion onto black holes have been developed:
\citet{2009_Booth_Schaye,2010_Booth_Schaye} assume that the accretion parameter $\alpha$ is not constant, but a function of the gas density, whereas in the EAGLE simulations \citep{2015_Schaye_Crain_Bower} and \citet{2015_Rosas-Guevara_Bower_Schaye} $\alpha$ is the ratio of Bondi time-scale and viscous time-scale.
\citet{2017_Tremmel_Karcher_Governato} modify the Bondi-Hoyle-Lyttleton parametrization to take the angular momentum of the gas into account.
\citet{2011_Debuhr_Quataert_Ma,2012_Debuhr_Quataert_Ma} calculate the black hole accretion rate based on the viscous evolution of an accretion disc that surrounds the black hole.
\citet{2011_Hopkins_Quataert} and the FIRE simulations \citep{2017_AnglesAlcazar_FaucherGiguere_Quataert} calculate the black hole accretion rate based on angular momentum and gravitational torques around the black hole.

\subsection{Black hole feedback}
Black hole accretion results in a luminosity of
\begin{equation}
  L = \epsilon_{\rmn{r}} \dot{M}_{\rmn{BH}} c^2 
\end{equation}
with speed of light $c$, and we assume that a fraction\footnote{Common values for the feedback efficiency are $\epsilon_{\rmn{f}} = 0.05$ \citep{2005_DiMatteo_Springel_Hernquist, 2005_Springel_Di-Matteo_Hernquist, 2007_Sijacki_Springel_DiMatteo, 2008_DiMatteo_Colberg_Springel, 2009_Johansson_Naab_Burkert}, $\epsilon_{\rmn{f}} = 0.15$ \citep{2009_Booth_Schaye, 2010_Booth_Schaye, 2015_Schaye_Crain_Bower} and $\epsilon_{\rmn{f}} = 0.02$ \citep{2017_Tremmel_Karcher_Governato}.} $\epsilon_{\rmn{f}} = 0.05$ of this luminosity is available as thermal energy for the gas that surrounds the black hole.
Thus the gas receives an energy per time of
\begin{equation}
  \dot{E} = \epsilon_{\rmn{f}} \epsilon_{\rmn{r}} \dot{M}_{\rmn{BH}} c^2
\end{equation}
that is distributed kernel weighted among the 50 nearest gas particles.
To avoid too large sound speeds and thus too small time steps we limit the specific energy of a single gas particle to $(0.1 c)^2$.

There are a few alternative models for black hole feedback.
\citet{2007_Sijacki_Springel_DiMatteo} use an additional feedback mode operating at low black hole accretion rates that is modeled by injecting \lq bubbles\rq\, into the host galaxy.
\citet{2011_Debuhr_Quataert_Ma,2012_Debuhr_Quataert_Ma} and \citet{2012_Choi_Ostriker_Naab} use \lq kinetic feedback\rq\, that feeds momentum and mass to the gas surrounding the black hole.

%

\section{Initial conditions, galaxy properties and parameters}\label{sec:ics}
Our galaxies and their properties are listed in Table \ref{tab:galaxies} at the end of this paper.
We take five galaxies from the original NIHAO suite that have a dark matter particle mass of $1.74 \times 10^6 \, \mathrm{M}_{\sun}$, a gas particle mass of $3.17 \times 10^5 \, \mathrm{M}_{\sun}$, a dark matter softening length of $931 \, \mathrm{pc}$ and a gas particle softening length of $398 \, \mathrm{pc}$.
These originate from a cosmological dark-matter-only simulation with a box size of 60 Mpc and 600$^3$ particles.
From the same box we additionally take seven new galaxies with the same resolution that reach $z=0$ halo masses of $1-4 \times 10^{12} \, \mathrm{M}_{\sun}$.
We then take 38 new galaxies with a dark matter particle mass of $1.38 \times 10^7 \, \mathrm{M}_{\sun}$, a gas particle mass of $2.52 \times 10^6 \, \mathrm{M}_{\sun}$, a dark matter softening length of $1863 \, \mathrm{pc}$ and a gas particle softening length of $782 \, \mathrm{pc}$.
These reach $z=0$ halo masses of $4 \times 10^{12}$ to $4 \times 10^{13} \, \mathrm{M}_{\sun}$, and originate from a cosmological dark-matter-only simulation with a box size of 90 Mpc and 450$^3$ particles.
A few galaxies were resimulated without black hole feedback, these have no entry for the black hole mass in Table \ref{tab:galaxies}.

The name of a galaxy refers to its halo mass at $z=0$ from the cosmological dark-matter-only simulations, i.e. its halo mass in the simulations presented in this paper might be slightly different.
We use the  AMIGA Halo Finder \citep[AHF,][]{2004_Gill_Knebe_Gibson,2009_Knollmann_Knebe} to identify haloes and their properties. 
AHF identifies overdensities on an adaptively refined grid as prospective halo centres.
These haloes are then defined as a sphere with radius $R_{{200}}$ and mass $M_{{200}}$ such that their density equals 200 times the cosmic critical matter density. 
The stellar mass $M_{\star}$ of a galaxy is defined as the combined mass of all stellar particles within a radius of $r_{\rmn{gal}} = 0.2 \, R_{{200}}$ 
\citep[see][for a different approach in defining a galaxy's stellar mass]{2013_Munshi_Governato_Brooks}.
%
Due to mergers some galaxies might contain more than one black hole, we define the central black hole as the one that is closest to the galaxy's centre.
All our haloes contain no intruder within 20 per cent of their virial radius and less than 20 intruders in total, where an intruder is a particle that originates from outside the zoom-in region.
In total we have 50 galaxies with black hole feedback and 11 galaxies without black hole feedback in this paper.

\section{Results}\label{sec:results}

The magnitude of black hole feedback is determined by the black hole accretion rate, which in turn determines the black hole mass.
As black hole feedback quenches star formation, one of its strongest effects is to reduce the stellar mass of a galaxy.
Therefore we use the galaxy's stellar mass and black hole mass to gauge our model, and thus use the stellar mass versus halo mass ($M_{\star}$-$M_{{200}}$) relation and the black hole mass versus stellar mass ($M_{\rmn{BH}}$-$M_{\star}$) relation for calibration.
We also analyze the star formation histories of three of our new galaxies to confirm black hole feedback having a quenching effect on them.
In this section we use the reference parameters of our model, namely the accretion parameter $\alpha = 70$, the feedback efficiency $\epsilon_{\rmn{f}} = 0.05$, the black hole seed mass $M_{\rmn{BH,s}} = 1 \times 10^5 \, \rmn{M}_{\sun}$ and the halo threshold mass $M_{\rmn{h,t}} = 5 \times 10^{10} \, \rmn{M}_{\sun}$.

We compare our $M_{\star}$-$M_{{200}}$ relation to results from halo abundance matching of
\citet{2018_Moster_Naab_White}, \citet{2013_Behroozi_Wechsler_Conroy} and \citet{2013_Moster_Naab_White}.
These are based on the IMF of \citet{2003_Chabrier}, but there is evidence for \lq heavier\rq\, (bottom-heavy, i.e. more low mass stars) IMFs in massive galaxies \citep[e.g.,][]{2012_Conroy_vanDokkum,2013_Dutton_Treu_Brewer,2013_Dutton_Maccio_Mendel}.
Therefore, following \citet{2013_Dutton_Maccio_Mendel}, we add a correction
\begin{equation}
\Delta_{\rmn{IMF}} = \min \left( 0.28 + \left( 0.14 m - 11 \right), 0.3 \right)
\end{equation}
for $m>9$ and $m = \log \left( M_{\star}/\rmn{M}_{\sun} \right)$ to $\log M_{\star}$ of the $M_{\star}$-$M_{{200}}$
relations, which increases the stellar mass for high mass galaxies by a factor of $\sim$2.
The \cite{2013_Behroozi_Wechsler_Conroy} and \citet{2018_Moster_Naab_White} relations are a function of the halo's virial mass, which we convert to $M_{200}$ by applying corrections suggested by \citet{2014_Dutton_Maccio}.

\begin{figure*}
  \includegraphics[width=170mm]{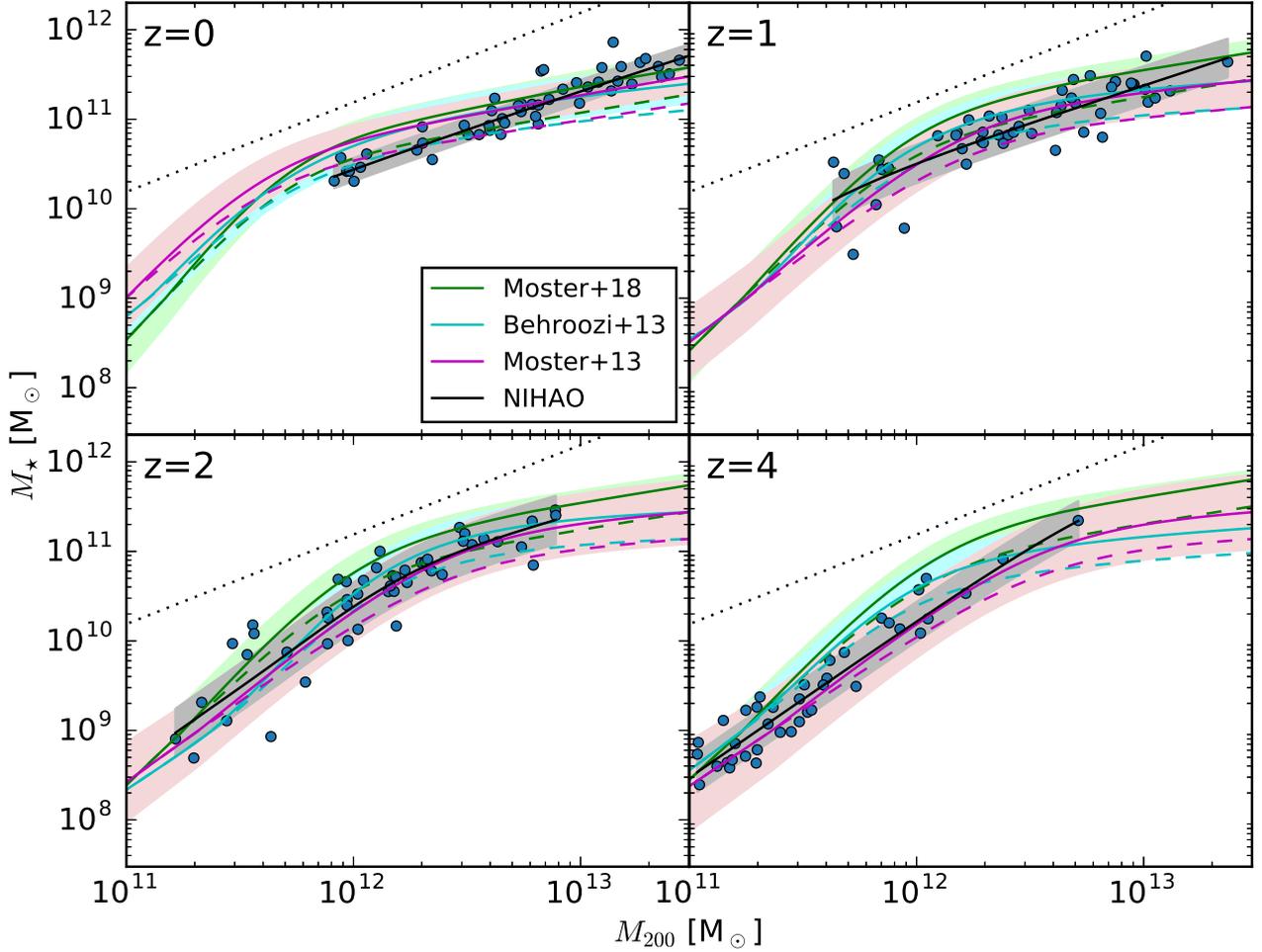}
  \caption{Stellar mass versus halo mass relation for our simulations (circles) for different redshifts.
                The solid lines show results from halo abundance matching of \citet{2018_Moster_Naab_White}, \citet{2013_Behroozi_Wechsler_Conroy}
                and \citet{2013_Moster_Naab_White} with IMF and halo mass corrections,
                and a fit of our NIHAO galaxies to eq.\,3 of \citet{2013_Behroozi_Wechsler_Conroy}. The shaded regions are the 1-sigma scatter.
                The dashed lines are the uncorrected \citet{2013_Moster_Naab_White} relation and the \citet{2018_Moster_Naab_White}
                and \citet{2013_Behroozi_Wechsler_Conroy} relations with corrections for the halo mass only.
                The dotted line is the cosmic baryon fraction.
                }
  \label{fig:mstar_mhalo}
\end{figure*}

\begin{figure}
  \includegraphics[width=84mm]{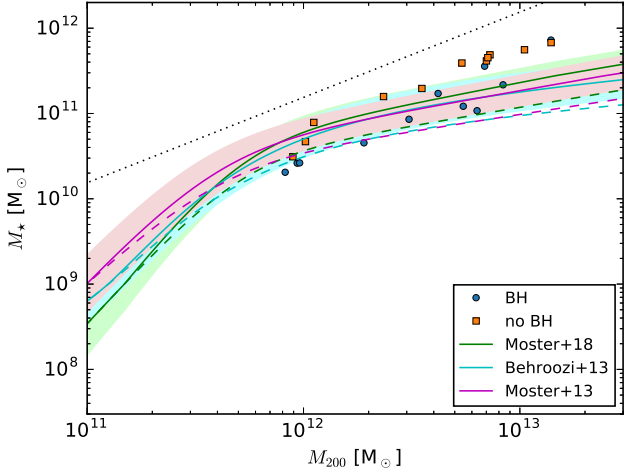}
  \caption{Stellar mass versus halo mass relation for our simulations for $z=0$,
                circles are galaxies with black holes, squares are galaxies without black holes.
                The solid lines show results from halo abundance matching of \citet{2018_Moster_Naab_White}, \citet{2013_Behroozi_Wechsler_Conroy}
                and \citet{2013_Moster_Naab_White} with IMF and halo mass corrections. The shaded regions are the 1-sigma scatter.
                The dashed lines are the uncorrected \citet{2013_Moster_Naab_White} relation and the \citet{2018_Moster_Naab_White}
                and \citet{2013_Behroozi_Wechsler_Conroy} relations with corrections for the halo mass only.                
                The dotted line is the cosmic baryon fraction.
                }
  \label{fig:mstar_mhalo_wobh}
\end{figure}

\begin{figure*}
  \includegraphics[width=170mm]{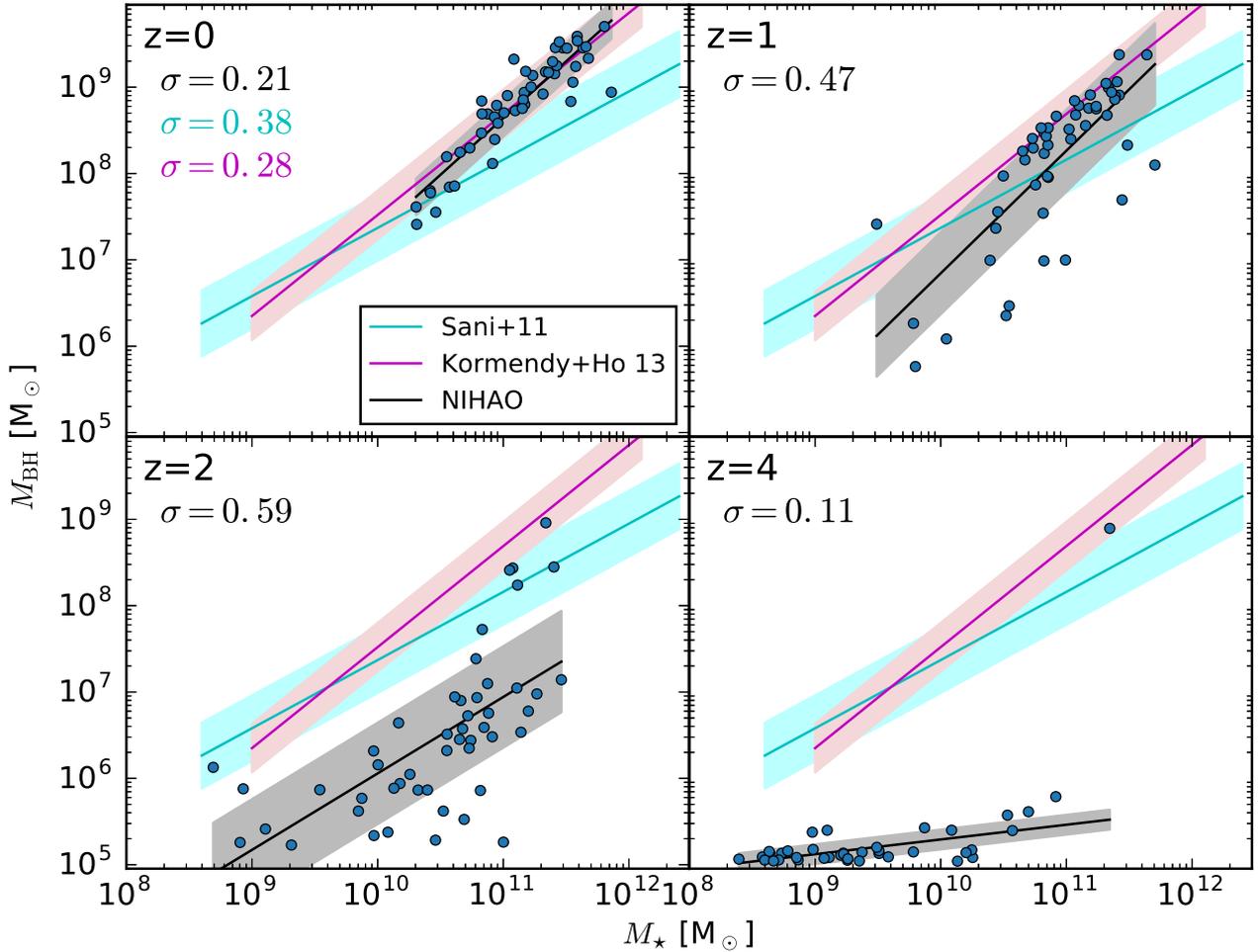}
  \caption{Black hole mass versus stellar mass relation for our simulations (circles) for different redshifts. The solid lines show
               observed $z=0$ relations from \citet{2011_Sani_Marconi_Hunt} and \citet{2013_Kormendy_Ho},
               and a linear fit to our NIHAO galaxies. The shaded regions are the 1-sigma scatter.
               }
  \label{fig:mbh_mstar}
\end{figure*}

\begin{figure}
  \includegraphics[width=84mm]{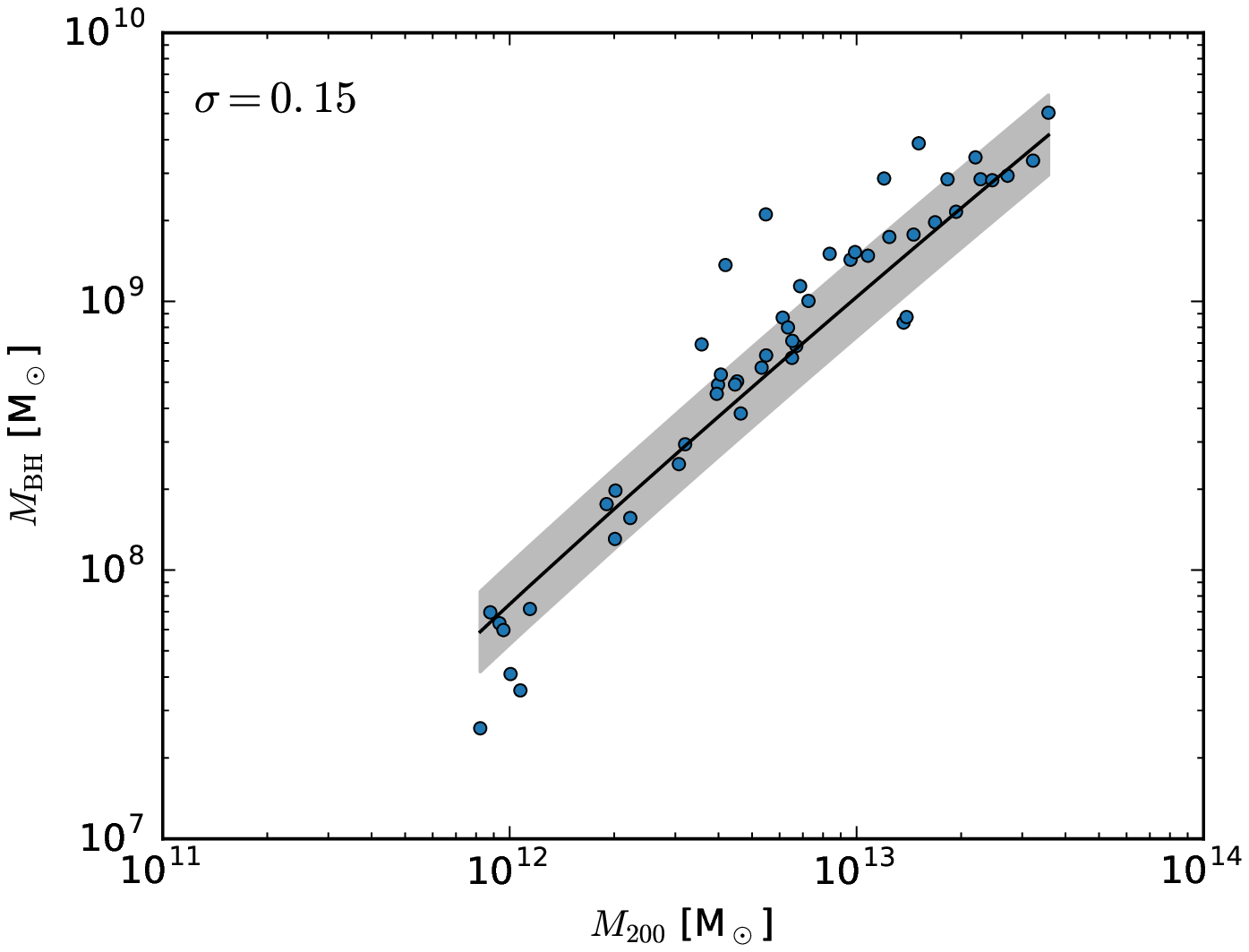}
  \caption{Black hole mass versus halo mass relation for our simulations (circles). The solid line is a combination of our $z=0$ fits
               to the $M_{\star}$-$M_{{200}}$ and the $M_{\rmn{BH}}$-$M_{\star}$ relations.
               The shaded region is the 1-sigma scatter.
               }
  \label{fig:mbh_mhalo}
\end{figure}

Fig. \ref{fig:mstar_mhalo} shows the $M_{\star}$-$M_{{200}}$ relation of our galaxies compared to the abundance matching results,
our simulations with black hole feedback provide a good match.
Only at very high and low halo masses our galaxies slightly deviate from the observed relations,
but most of them are still within the 1-sigma scatter of the observations.
Fig. \ref{fig:mstar_mhalo_wobh} shows the $M_{\star}$-$M_{{200}}$ relation of our galaxies that were simulated
without black holes, compared to their counterpart with black holes.
Only the simulations with black holes provide a good match to the observed relations,
demonstrating the importance of black hole feedback for the evolution of high mass galaxies.
Fig. \ref{fig:mbh_mstar} shows the $M_{\rmn{BH}}$-$M_{\star}$ relation of our galaxies for different redshifts compared with $z=0$ observations from \citet{2013_Kormendy_Ho} and \citet{2011_Sani_Marconi_Hunt}.
At each redshift we fit a linear relation to our simulations and calculate the 1-sigma scatter.
We apply a 5-sigma clipping to our data to remove the high-mass outlier that is visible in the $z=4$ subplot from the calculation of the scatter.
Our results match the \citet{2013_Kormendy_Ho} relation well,
only at the lower end the black hole masses of some galaxies are underestimated.
We also show a relation between halo mass and black hole mass in Fig. \ref{fig:mbh_mhalo}.

Recent successes in reproducing the correct stellar masses and black hole masses in galaxy simulations include the EAGLE project \citep{2015_Schaye_Crain_Bower}, which consists of cosmological simulations that reproduce the $M_{\star}$-$M_{200}$ relation in $\rmn{log}\,M_{200}$ from 10.5 to 14.5 and the $M_{\rmn{BH}}$-$M_{\star}$  relation in $\rmn{log}\,M_{\star}$ from 8 to 12.
The $M_{\star}$-$M_{200}$ relation of the Illustris project \citep{2014_Genel_Vogelsberger_Springel,2014_Vogelsberger_Genel_Springel} is in good qualitative agreement with observations in $\rmn{log}\,M_{200}$ from 10 to 14, also their $M_{\rmn{BH}}$-$M_{\rmn{bulge}}$ relation (in $\rmn{log}\,M_{\rmn{bulge}}$ from 8 to 12) is in good agreement with observations \citep{2015_Sijacki_Vogelsberger_Genel}.
The IllustrisTNG simulations \citep{2018_Weinberger_Springel_Pakmor} reproduce the $M_{\rmn{BH}}$-$M_{\star}$ relation in $\rmn{log}\,M_{\star}$ from 8 to 12.
Also The ROMULUS cosmological simulations \citep{2017_Tremmel_Karcher_Governato}
match the $M_{\star}$-$M_{{\rmn{vir}}}$ relation in $\rmn{log}\,M_{\rmn{vir}}$ from 10 to 13
and the $M_{\rmn{BH}}$-$M_{\rmn{\star}}$ relation in $\rmn{log}\,M_{\rmn{\star}}$ from 9 to 11.5.
The Magneticum Pathfinder simulations \citep{2014_Hirschmann_Dolag_Saro} match the 
$M_{\rmn{BH}}$-$M_{\rmn{stellar}}$ relation in $\rmn{log}\,M_{\rmn{stellar}}$ from 10.5 to 13.

\begin{figure}
  \includegraphics[width=84mm]{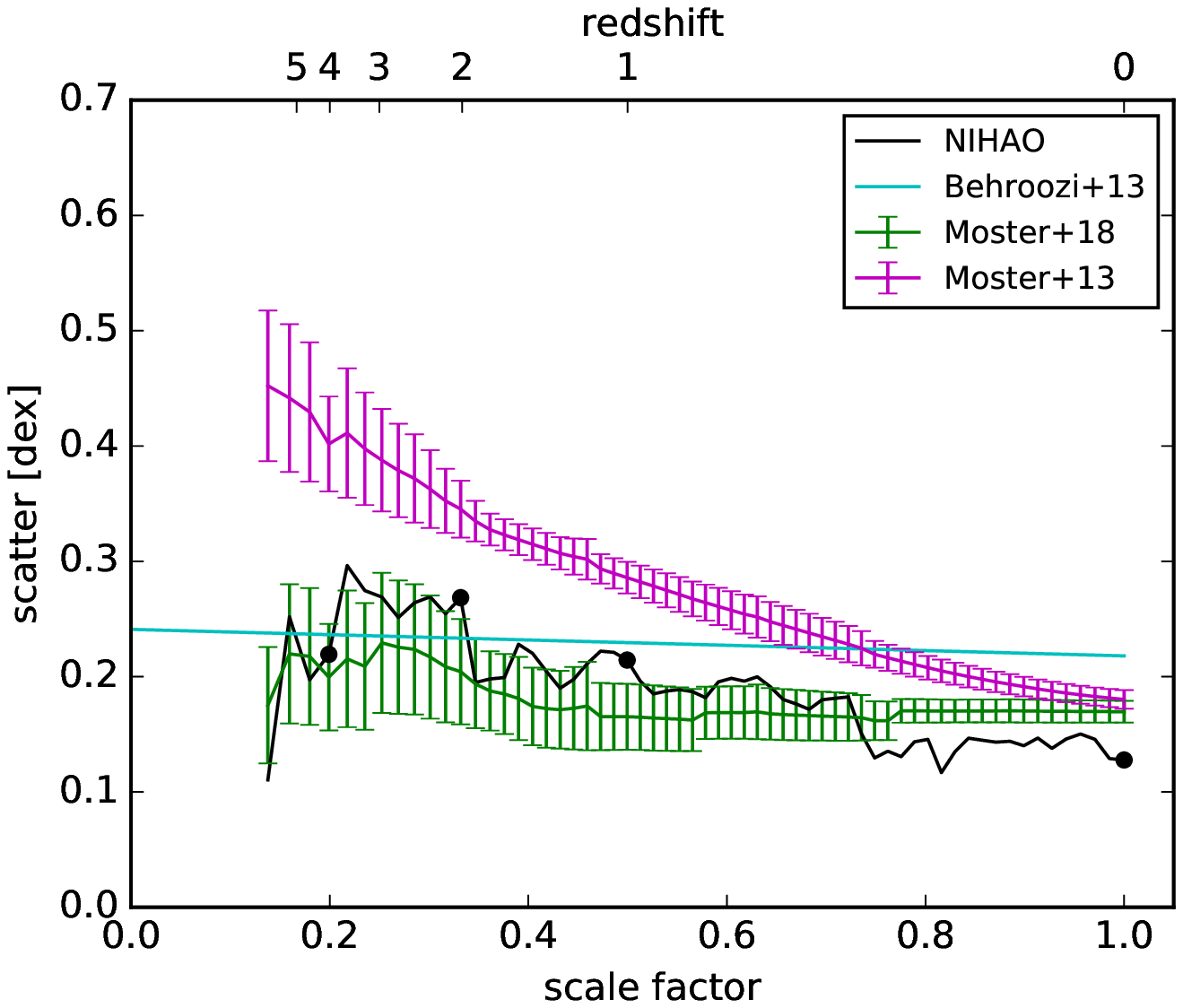}
  \caption{Scatter of the stellar mass versus halo mass relation versus time for the \citet{2018_Moster_Naab_White},
                \citet{2013_Behroozi_Wechsler_Conroy} and \citet{2013_Moster_Naab_White}
                stellar mass halo mass functions and for our NIHAO galaxies.                
                The black circles mark the redshifts shown in Fig. \ref{fig:mstar_mhalo}.
                }
  \label{fig:mstar_mhalo_scatter}
\end{figure}

Fig. \ref{fig:mstar_mhalo_scatter} shows the scatter of the $M_{\star}$-$M_{{200}}$ relation versus time
for our simulations and the \citet{2018_Moster_Naab_White}, \citet{2013_Behroozi_Wechsler_Conroy} and \citet{2013_Moster_Naab_White} relations.
The scatter of the Moster relations depends on the stellar mass,
therefore we calculate the scatter for the range of stellar masses
that are encompassed by our simulations in each redshift, its mean and standard deviation is shown in Fig. \ref{fig:mstar_mhalo_scatter}.
We make sure that at each redshift there are at least 12 galaxies in the sample given our criteria of a minimum of $10^4$ particles per galaxy.
The scatter of all relations is generally decreasing with time.
Only at higher redshifts the scatter of our relation is rising,
possibly caused by low number statistics.
The scatter of our relation is always below the \citet{2013_Moster_Naab_White} scatter,
and below the \citet{2013_Behroozi_Wechsler_Conroy} scatter for z<1.5.
The \citet{2018_Moster_Naab_White} scatter is in good agreement with our simulations,
with being only slightly below our scatter for z>1 and above for z<0.5.

\begin{figure}
  \includegraphics[width=84mm]{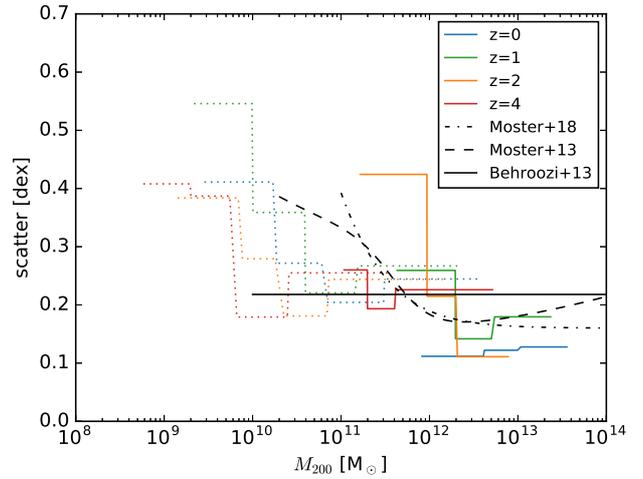}
  \caption{Scatter of the stellar mass versus halo mass relation as a function of halo mass for the $z=0$ 
                \citet{2018_Moster_Naab_White}, \citet{2013_Behroozi_Wechsler_Conroy} and \citet{2013_Moster_Naab_White}
                stellar mass halo mass functions, and, for four different redshifts, for the 50 NIHAO galaxies with black holes presented in this paper (solid lines)
                and for 78 of the NIHAO galaxies without black holes presented in \citet{2015_Wang_Dutton_Stinson} (dotted lines).}
  \label{fig:mstar_mhalo_scatter_mass}
\end{figure}

Fig. \ref{fig:mstar_mhalo_scatter_mass} shows the scatter of the $M_{\star}$-$M_{{200}}$ relation versus halo mass
for $z=0$ for the \citet{2013_Behroozi_Wechsler_Conroy}, \citet{2018_Moster_Naab_White} and \citet{2013_Moster_Naab_White}
relations, for the NIHAO galaxies with black holes presented in this paper (50 galaxies divided into 3 bins) and for 78 of the NIHAO
galaxies (divided into 4 bins) without black holes presented in \citet{2015_Wang_Dutton_Stinson} for four different redshifts.
For $z=0$ our scatter is below the \citet{2018_Moster_Naab_White} and \citet{2013_Moster_Naab_White} scatter,
except for halo masses around $10^{10} \, \mathrm{M}_{\sun}$, and below the \citet{2013_Behroozi_Wechsler_Conroy}
scatter for halo masses above $5 \times 10^{11} \, \mathrm{M}_{\sun}$.
The scatter is generally increasing with decreasing halo mass.
There is no clear trend as a function of redshift.
Our results are in agreement with previous studies of the scatter by \citet{2018_Wechsler_Tinker} and \citet{2017_Matthee_Schaye_Crain}.
So far only the NIHAO simulations can show the scatter for halo masses covering five orders of magnitude, i.e., ranging from $5 \times 10^{8}$ to $4 \times 10^{13} \, \mathrm{M}_{\sun}$.

\begin{figure}
  \includegraphics[width=84mm]{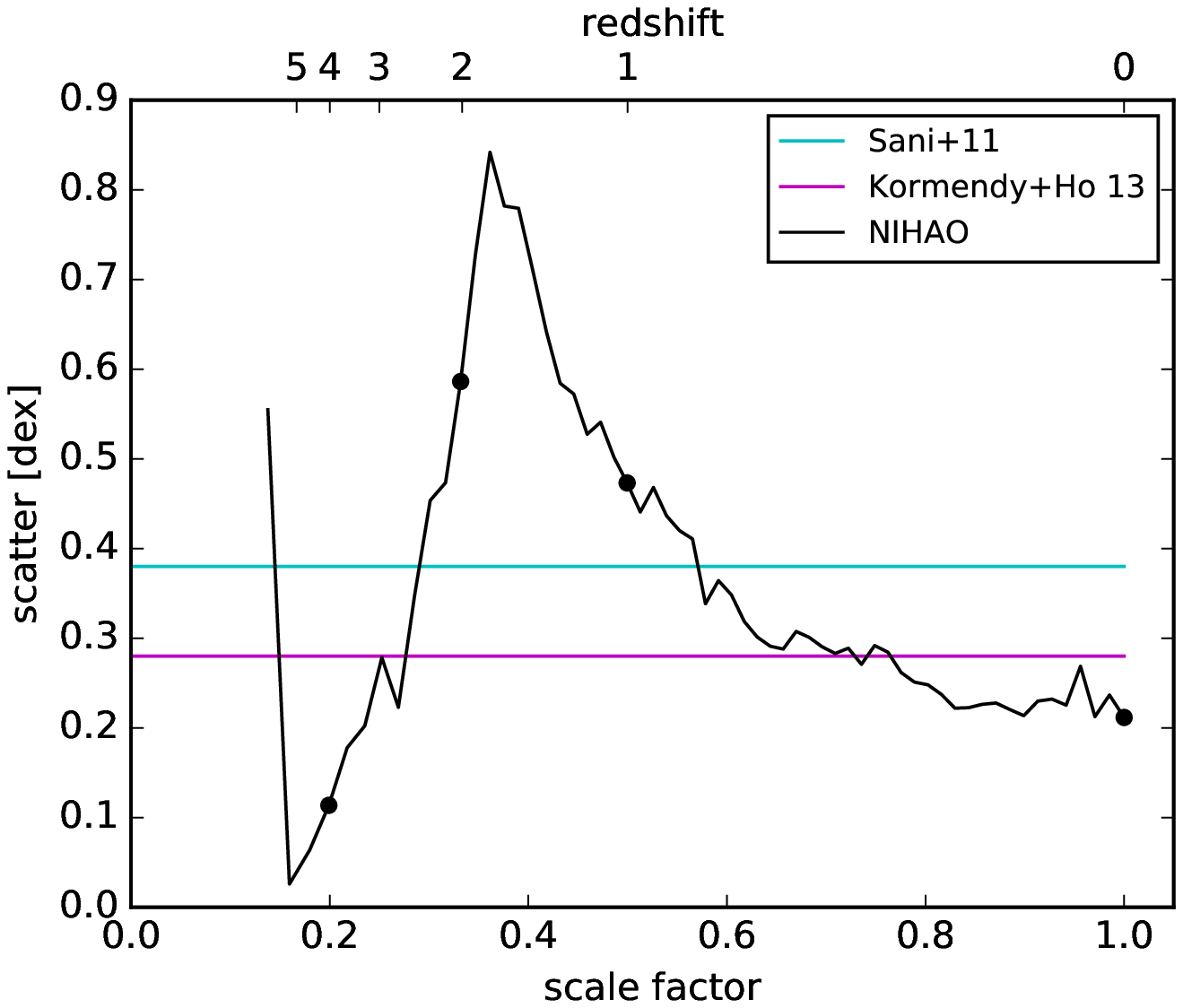}
  \caption{Scatter of the black hole mass versus stellar mass relation versus time for the
               observed $z=0$ relations from \citet{2011_Sani_Marconi_Hunt}
               and \citet{2013_Kormendy_Ho} and for our NIHAO galaxies.
               The black circles mark the redshifts shown in Fig. \ref{fig:mbh_mstar}.
               }
  \label{fig:mbh_mstar_scatter}
\end{figure}

Fig. \ref{fig:mbh_mstar_scatter} shows the scatter of the $M_{\rmn{BH}}$-$M_{\star}$ relation versus time
for our simulations and the $z=0$ scatter of the \citet{2011_Sani_Marconi_Hunt} and \citet{2013_Kormendy_Ho} relations.
We only calculate the scatter for redshifts with at least 12 galaxies in our sample.
At high redshifts the scatter is dominated by the black hole seed mass and possibly by low number statistics.
The scatter then increases steeply, and is then declining for the remainder of the evolution.
At redshift zero the scatter of the simulations is below the observed scatter.
This is expected since we present here the intrinsic scatter in the simulations which does not account for observational biases and errors.

\begin{figure*}
  \includegraphics[width=177mm]{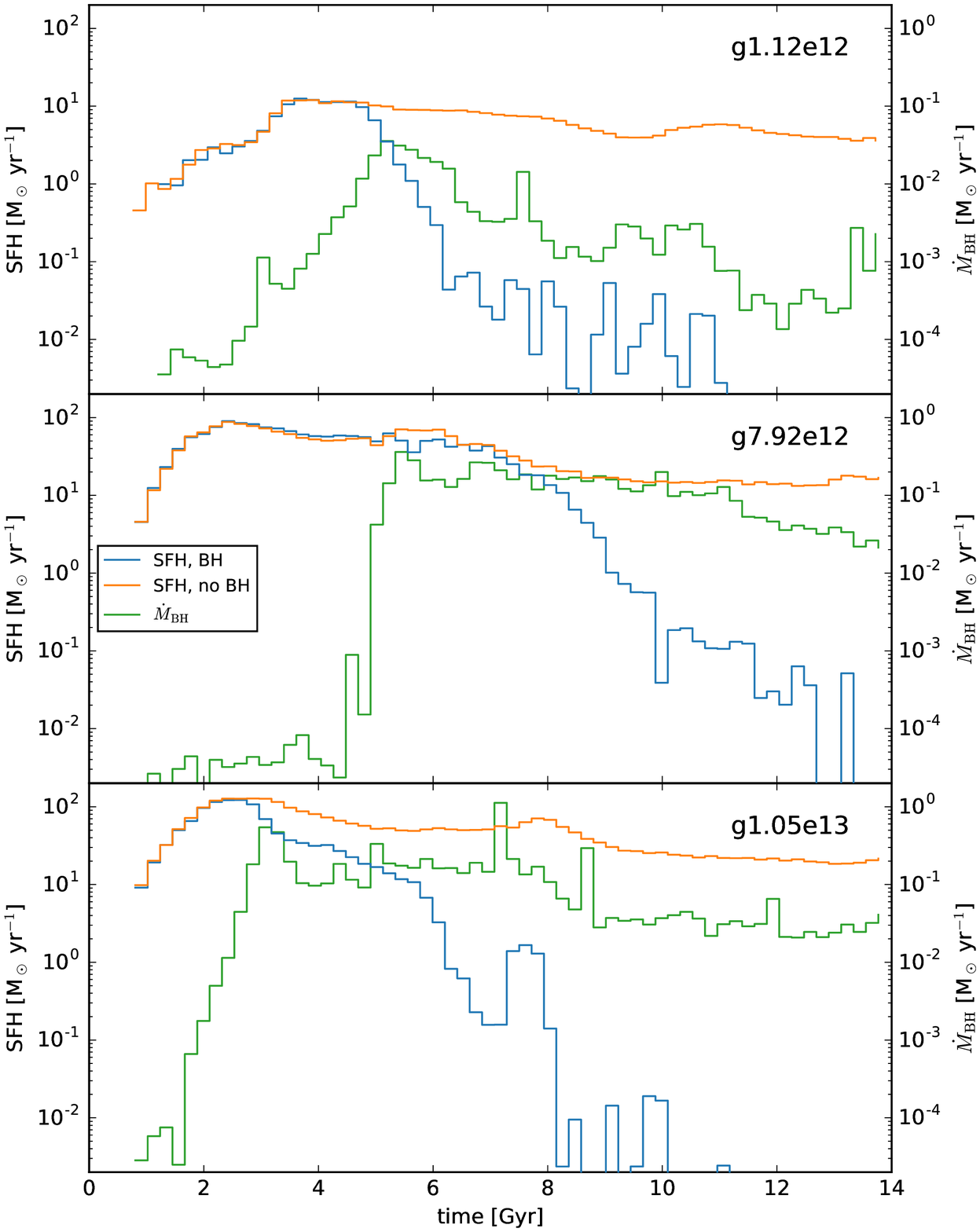}
  \caption{Star formation history, with and without black holes (BH), and the black hole
                accretion rate for the elliptical galaxies g1.12e12, g7.92e12 and g1.05e13.
                The time binning is constant and equal to 216 Myr.}
  \label{fig:sfh}
\end{figure*}

In Fig. \ref{fig:sfh} we show the star formation history, with and without black hole feedback, and the black hole accretion rate of three of our galaxies.
Without feedback the galaxies show continuous star formation up to $z=0$, creating unrealistic and unobserved blue massive galaxies.
With feedback the galaxies show an initial phase of high star formation, and are then quenched for the remainder of their evolution.
The back hole accretion rate peaks shortly before the star formation rate starts declining, indicating that star formation is quenched by the black hole feedback.

\begin{figure*}
  \includegraphics[width=162mm]{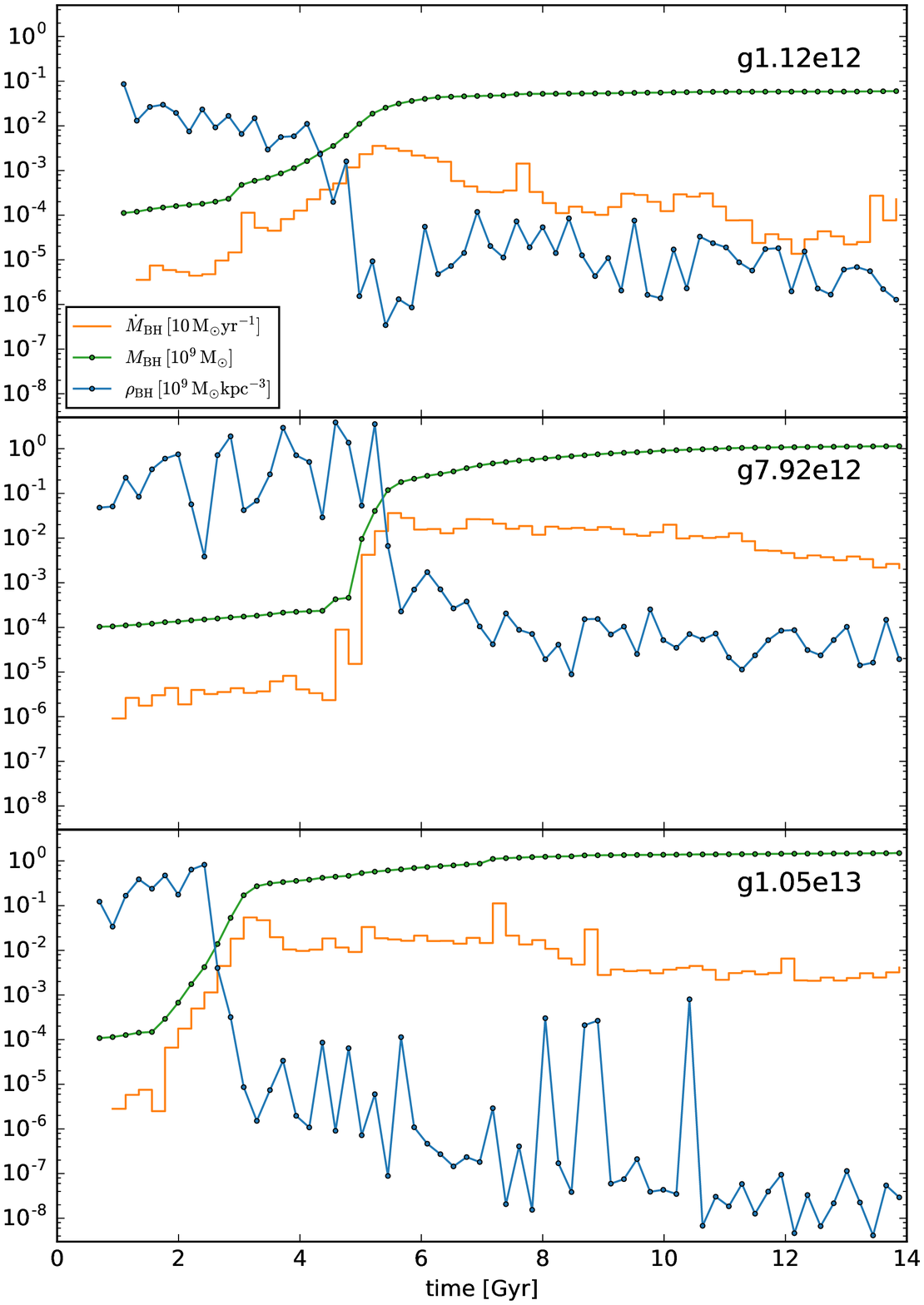}
  \caption{Black hole mass, black hole accretion rate and SPH density around the black hole
                for the elliptical galaxies g1.12e12, g7.92e12 and g1.05e13.
                The time binning is constant and equal to 216 Myr.}
  \label{fig:time_bh}
\end{figure*}

All three galaxies in Fig. \ref{fig:sfh} show the black hole becoming active at about 2 to 4 gigayears.
To determine the reason for this increase we look at Fig. \ref{fig:time_bh}, which shows the black hole accretion rate,
the black hole mass and the gas density at the location of the black hole.
The black hole accretion rate is given by the Bondi formula of eq. \ref{eq:bondi}, and is determined mainly by the black hole mass and the gas density.
In the first few time steps both quantities results in a low growth rate for all three galaxies, which would, if continuing until redshift zero,
result in final black hole masses of less than $10^{7} \, \rmn{M}_{\sun}$.

However, after the first few time steps
(9 steps or 2.8 Gyr for g1.12e12, 18 steps or 4.3 Gyr for g7.92e12, 5 steps or 1.5 Gyr for g1.05e13)
the central black hole experiences one or more merger events with other black holes. Thus the black hole mass roughly doubles, which roughly quadruples the black hole accretion rate. Now the black hole accretion rate is large enough to cause significant black hole growth, and the ever increasing black hole mass leads to an ever increasing accretion rate, resulting in a kind of \lq runaway growth\rq.
The black hole gains most of its mass due to gas accretion, the black hole mergers serve merely as a trigger for higher accretion rates.
For instance, the final central black hole mass of the galaxy g7.92e12 consists to 5 per cent of merged black holes and to 95 per cent of accreted gas.

When the black hole accretion rate, and thus the feedback, is high enough, it is capable of removing significant amounts of gas from the vicinity of the black hole, indicated by a drop in the gas density in Fig. \ref{fig:time_bh}. An increasing black hole mass and a decreasing gas density leads the black hole accretion rate to stabilize to a roughly constant value, at least for the next few gigayears.


Previous work \citep{2017_AnglesAlcazar_FaucherGiguere_Quataert,2015_Dubois_Volonteri_Silk} indicates that black hole growth is regulated by the gas inflow from the host galaxy, and that black hole growth is only efficient after the galaxy has reached a certain mass.
However, in our work the gas density does not change significantly in the first stages of black hole growth,
rather, black hole mergers are crucial in triggering significant black hole growth.
This implies the existence of a critical black hole mass for the onset of black hole accretion of the order of  $10^{5} \, \rmn{M}_{\sun}$.
Furthermore it outlines the importance of triggering efficient black hole growth in the first place, alongside with self-regulated black hole growth in the later stages of galaxy evolution.
However, we point out that these effects are merely the result of the Bondi formula
being proportional to the square of the black hole mass, and might change if other accretion schemes are used.

\section{Parameter study}\label{sec:paramstudy}
The reference parameters of our model are the accretion parameter $\alpha = 70$, the feedback efficiency $\epsilon_{\rmn{f}} = 0.05$, the black hole seed mass $M_{\rmn{BH,s}} = 1 \times 10^5 \, \rmn{M}_{\sun}$ and the halo threshold mass $M_{\rmn{h,t}} = 5 \times 10^{10} \, \rmn{M}_{\sun}$.
In this section we explore how a variation of these model parameters affect the $M_{\star}$-$M_{{200}}$ and $M_{\rmn{BH}}$-$M_{\star}$ relations, and show that our fiducial parameters are a reasonable choice. We restrict the parameter study to the galaxies g8.26e11 and g7.92e12,
the first one is the same galaxy used in the first NIHAO paper \citet{2015_Wang_Dutton_Stinson} to calibrate the stellar feedback, while the second one has been chosen since it is one order of magnitude larger in halo mass. 

In this parameter study we vary only one parameter at the same time, thus it is possible that changing multiple parameters simultaneously gives an even better match to the observed scaling relations.
Further improvement could be reached by varying parameters not associated to black hole feedback, e.g., parameters related to stellar feedback, or by changing the models for black hole formation, accretion and feedback as outlined in section \ref{sec:numerics}.

\begin{figure*}
  \includegraphics[width=162mm]{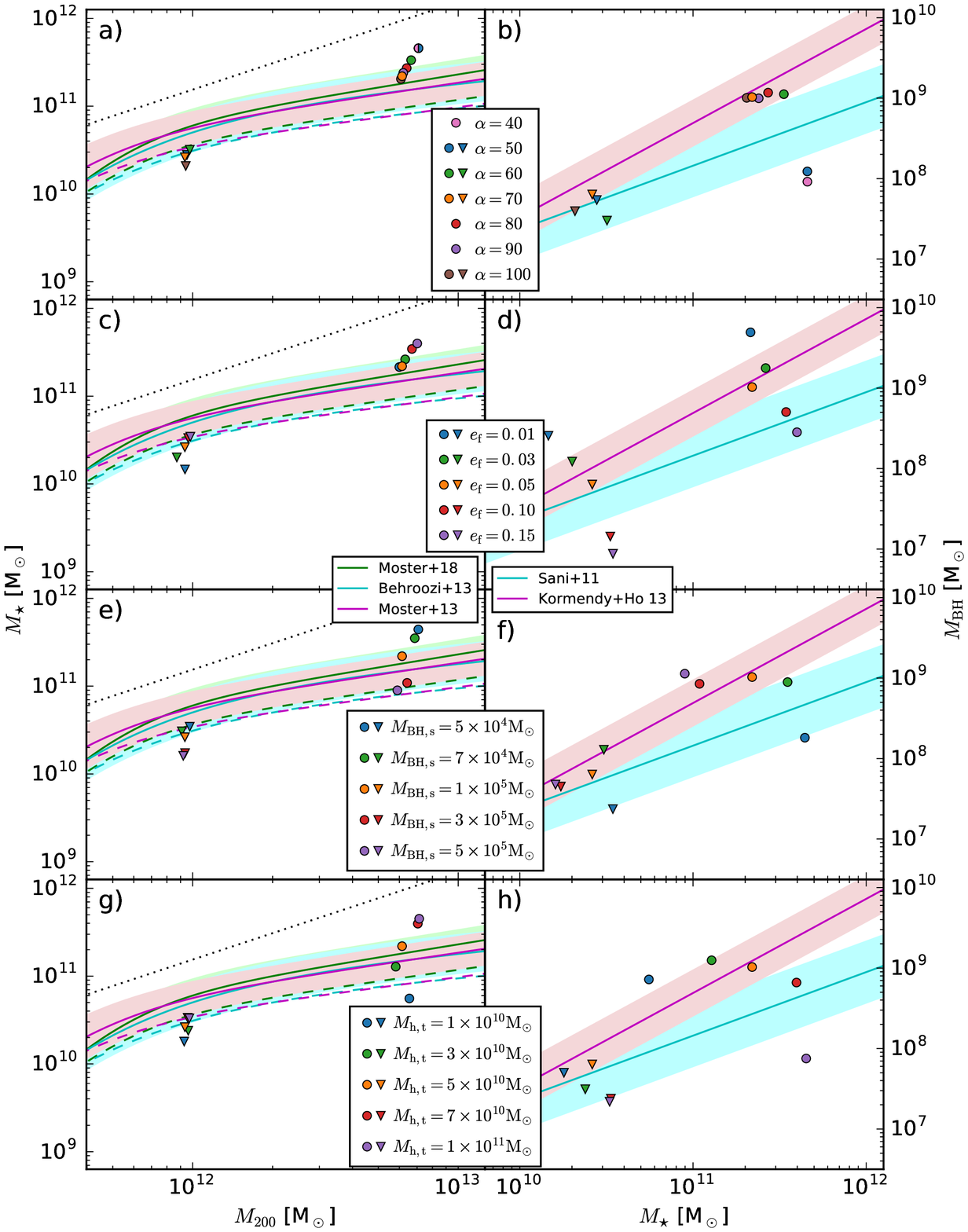}
  \caption{Parameter study for the galaxies g7.92e12 (circles) and g8.26e11 (triangles)
                for variations of the accretion parameter $\alpha$ (first row), feedback efficiency
                $\epsilon_{\rmn{f}}$ (second row), black hole seed mass $M_{\rmn{BH,s}}$ (third row)
                and halo threshold mass $M_{\rmn{h,t}}$ (fourth row).
                Left column: stellar mass versus halo mass relation.
                The solid lines show results from halo abundance matching of \citet{2018_Moster_Naab_White}, \citet{2013_Behroozi_Wechsler_Conroy}
                and \citet{2013_Moster_Naab_White} with IMF and halo mass corrections,
                and a fit of our NIHAO galaxies to eq.\,3 of \citet{2013_Behroozi_Wechsler_Conroy}. The shaded regions are the 1-sigma scatter.
                The dashed lines are the uncorrected \citet{2013_Moster_Naab_White} relation and the \citet{2018_Moster_Naab_White}
                and \citet{2013_Behroozi_Wechsler_Conroy} relations with corrections for the halo mass only.                
                The dotted line is the cosmic baryon fraction.
                Right column: black hole mass versus stellar mass relation.
                The solid lines and shaded regions are observed relations from \citet{2011_Sani_Marconi_Hunt} and \citet{2013_Kormendy_Ho}.}
  \label{fig:param}
\end{figure*}

\begin{figure}
  \includegraphics[width=84mm]{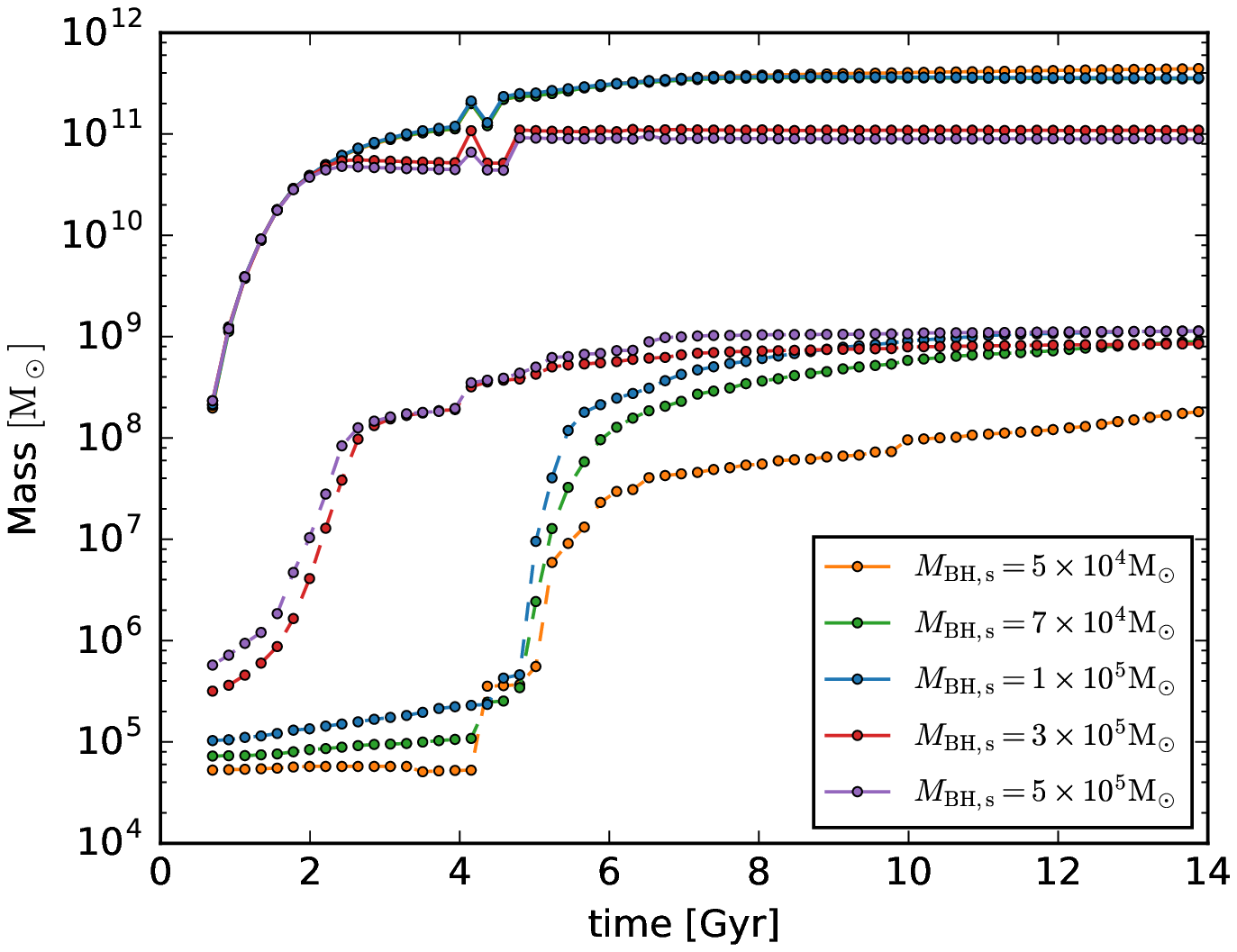}
  \caption{Black hole mass (lower dashed lines) and stellar mass (upper solid lines) versus time for the galaxy g7.92e12
                for different black hole seed masses.}
  \label{fig:param_masses}
\end{figure}

\subsection{Accretion parameter \texorpdfstring{$\alpha$}{Lg}}

For the galaxy g7.92e12 the accretion parameter $\alpha$ seems to have a threshold value of about $\alpha=60$.
For $\alpha>60$ the variation in stellar mass (Fig. \ref{fig:param}a) does not show a clear trend but is possibly of stochastic nature. The black hole mass does not change significantly (Fig. \ref{fig:param}b), demonstrating the self-regulating nature of black hole feedback.
However, for $\alpha<60$ the black hole feedback is insufficient to reduce the stellar mass to match the observed $M_{\star}$-$M_{{200}}$ relations (Fig. \ref{fig:param}a) and the black hole mass rapidly drops below the observed $M_{\rmn{BH}}$-$M_{\star}$ relations (Fig. \ref{fig:param}b).
For the galaxy g8.26e11 this threshold value seems to be smaller, possibly due to a higher resolution, as all simulations only show small variations in the $M_{\star}$-$M_{{200}}$ and $M_{\rmn{BH}}$-$M_{\star}$ relations that are possibly of stochastic nature.
Thus an accretion parameter of $\alpha=70$ seems to be a reasonable choice.

\subsection{Feedback efficiency \texorpdfstring{$e_{\rmn{f}}$}{Lg}}

A higher feedback efficiency quenches the accretion of gas by the black hole, and thus leads to a lower black hole mass (Fig. \ref{fig:param}d).
The total amount of feedback energy injected into the gas for the galaxy g7.92e12 is
$E = e_{\rmn{f}} e_{\rmn{r}} c^2 M_{\rmn{BH}} = (8.9, 9.6, 9.3, 9.1, 7.6) \times 10^{53} \rmn{J}$ for the values $e_{\rmn{f}} = (0.01, 0.03, 0.05, 0.10, 0.15)$, i.e. even changing the feedback efficiency by a factor of 15  changes the total feedback energy by less than 20 per cent.
Thus a higher feedback efficiency is compensated by a lower black hole mass, leading to approximately the same total feedback energy, again demonstrating the self-regulating nature of black hole feedback.
The stellar mass is slightly decreasing with decreasing feedback efficiency (Fig. \ref{fig:param}c), but this effect is small and could also be of stochastic nature.
Feedback efficiencies of 0.10 and 0.15 far overpredict the $M_{\star}$-$M_{{200}}$ relation for the galaxy g7.92e12 (Fig. \ref{fig:param}c), and feedback efficiencies of 0.01 and 0.03 far underpredict the $M_{\rmn{BH}}$-$M_{\star}$ relation for the galaxy g8.26e11 (Fig. \ref{fig:param}d), making a feedback efficiency of $e_{\rmn{f}}=0.05$ an optimal choice.

\subsection{Black hole seed mass \texorpdfstring{$M_{\rmn{BH,s}}$}{Lg}}

Small seed masses of $5 \times 10^4$ and $7 \times 10^4 \, \rmn{M}_{\sun}$ provide insufficient feedback, leading the galaxy g7.92e12 to highly exceed the observed $M_{\star}$-$M_{{200}}$ relations (Fig. \ref{fig:param}e).
High seed masses of $3 \times 10^5$ and $5 \times 10^5 \, \rmn{M}_{\sun}$ bring the galaxy g7.92e12 to the edges of the observed $M_{\star}$-$M_{{200}}$ and $M_{\rmn{BH}}$-$M_{\star}$ relations (Fig. \ref{fig:param}e,f), leaving a seed mass of $M_{\rmn{BH,s}}= 1 \times 10^5 \rmn{M}_{\sun}$ as optimal choice.

According to Fig. \ref{fig:param}e,f the final black hole mass does not depend on the black hole seed mass (except for the lowest seed mass), but the final stellar mass does. This can be explained with Fig. \ref{fig:param_masses}, which shows the black hole mass and the stellar mass as a function of time for the galaxy g7.92e12 and for different black hole seed masses.
The three lowest black hole seed masses show a black hole mass evolution as outlined in section \ref{sec:results}: The growth rate is low at first, then black hole mergers trigger \lq runaway growth\rq\, that is then stopped by feedback and ultimately leads to black hole masses in accordance with observations.

The two highest black hole seed masses show an entirely different evolution: Black hole mergers are not needed, the seed mass is already sufficiently high to trigger \lq runaway growth\rq.

Although the evolution of the black hole masses in both scenarios is entirely different at early times, they produce the same final black hole masses (except for the lowest seed mass), which can be attributed to the self-regulating nature of black hole feedback: Large black holes exert high feedback and allow for a low accretion rate, small black holes exert low feedback and allow for a high accretion rate.

Fig. \ref{fig:param_masses} also shows the stellar mass as a function of time for different initial black hole seed masses. Until 2.5 gigayears the stellar masses evolve almost identical, then they start diverging: The higher the black hole seed mass the earlier the black hole accretion rate peaks, thus star formation is quenched earlier, which sets the stellar masses to ever lower values. The sudden jumps in the stellar mass at about 4-5 gigayears are caused by a merger event.


Different black hole seed masses lead to the same black hole mass, but to different stellar masses at redshift zero. However, calibration on scaling relations can be used to determine the optimal choice for the black hole seed mass.

\subsection{Halo threshold mass \texorpdfstring{$M_{\rmn{h,t}}$}{Lg}}

High halo threshold masses of $7 \times 10^{10}$ or $7 \times 10^{10} \, \rmn{M}_{\sun}$
overpredict the $M_{\star}$-$M_{{200}}$ relation (Fig. \ref{fig:param}g)
for the galaxy g7.92e12,
whereas decreasing the halo threshold mass to $1 \times 10^{10}$ or $3 \times 10^{10} \, \rmn{M}_{\sun}$ moves the galaxy to the edge of the observed $M_{\rmn{BH}}$-$M_{\star}$ relations (Fig. \ref{fig:param}h).
Thus a halo threshold mass of $M_{\rmn{h,t}}=5 \times 10^{10} \rmn{M}_{\sun}$ is the most reasonable choice.

The results in Fig. \ref{fig:param}g,h resemble the results in Fig. \ref{fig:param}e,f:
Increasing the halo threshold mass is equivalent to reducing the black hole seed mass,
thus the same ratio of halo threshold mass to black hole seed mass produces the same results.
This can be explained as follows: The halo threshold mass basically determines the time the black hole is seeded.
If with our standard parameters a black hole with $10^{5} \, \rmn{M}_{\sun}$ is seeded at time t,
increasing the halo threshold mass means the black hole with $10^{5} \, \rmn{M}_{\sun}$ is seeded at a time t+dt,
and reducing the black hole seed mass means a black hole with $10^{4} \, \rmn{M}_{\sun}$ is seeded at time t.
The former is equivalent to the latter, as the black hole with $10^{4} \, \rmn{M}_{\sun}$ seeded at time t
will have grown to $10^{5} \, \rmn{M}_{\sun}$ at the time t+dt.

\section{Summary}\label{sec:summary}
We introduce and test algorithms for black hole formation, accretion and feedback to the NIHAO project. 
For black hole formation we place a black hole in the centre of a halo once it exceeds a threshold mass, for black hole accretion we use the Bondi-Hoyle-Lyttleton parametrization, and for black hole feedback we deposit thermal energy, which is proportional to the black hole accretion rate, into the gas that surrounds the black hole.
This addition to the NIHAO project allows us to extend the NIHAO suite of galaxies to higher masses.

Our galaxies show good agreement with the observed $M_{\star}$-$M_{{200}}$ and $M_{\rmn{BH}}$-$M_{\star}$ relations that we use to calibrate the free parameters of our model.
We also investigate the scatter of these relations and their time evolution. The scatter of both relations is decreasing with time for $z<1$ (higher redshifts possibly suffer from low number statistics and seeding effects), and is also lower than the observed scatter, possibly because we measure the intrinsic scatter of these relations without any observational uncertainties.

In the high mass, elliptical galaxies the quenching of star formation occurs after an increase of the black hole accretion rate, confirming that star formation is quenched by the black hole feedback. A parameter study confirms that we have chosen the optimal parameters within the framework of our model.
Our simulations provide a valuable tool to study the effect of black hole feedback on galaxy formation and evolution.


\section*{Acknowledgements}

The authors gratefully acknowledge the Gauss Centre for Supercomputing e.V. (www.gauss-centre.eu) for funding this project by providing computing time on the GCS Supercomputer SuperMUC at Leibniz Supercomputing Centre (www.lrz.de).
A part of this research was carried out on the High Performance Computing resources at New York University Abu Dhabi.
We used the software package {\sc pynbody} \citet{pynbody} for our analyses.
AO is funded by the Deutsche Forschungsgemeinschaft (DFG, German Research Foundation) - MO 2979/1-1.
We thank Tobias Buck for supplying the initial conditions for g7.92e12, g1.05e13 and g1.44e13.

\bibliographystyle{mnras}
\bibliography{library}


\begin{table*}
  \caption{Galaxies and their properties: particle number $N$, dark matter particle number $N_{\mathrm{DM}}$,
                star particle number $N_{\star}$, halo mass $M_{200}$, stellar mass $M_{\star}$
                and black hole mass $M_{\mathrm{BH}}$.
                First section: galaxies from the original NIHAO sample from the 60 Mpc box,
                second section: new galaxies from the 60 Mpc box,
                third section: new galaxies from the 90 Mpc box.
                Galaxies without black hole mass are re-simulations without black holes.
  }
  \label{tab:galaxies}
  \begin{tabular}{c|c|c|c|c|c|c}
      galaxy         &  $N$  &  $N_{\mathrm{DM}}$  &  $N_{\star}$  &  $\log \left( M_{200} [\mathrm{M}_{\sun}] \right)$  &  $\log \left( M_{\star} [\mathrm{M}_{\sun}] \right)$  &  $\log \left( M_{\mathrm{BH}} [\mathrm{M}_{\mathrm{\sun}}] \right)$    \\\hline
g7.55e11  &   932961  &   431968  &   327418  &  11.92  &  10.31  &   7.41  \\
g8.26e11  &  1070666  &   496667  &   424753  &  11.97  &  10.42  &   7.80  \\
g1.12e12  &  1062175  &   516394  &   430205  &  11.98  &  10.42  &   7.78  \\
g1.92e12  &  1960845  &  1035887  &   736102  &  12.28  &  10.66  &   8.25  \\
g2.79e12  &  3383907  &  1663073  &  1395558  &  12.49  &  10.93  &   8.39  \\
\hline
g7.55e11  &  1185149  &   455930  &   483752  &  11.95  &  10.49  &    -  \\
g8.26e11  &  1513265  &   518236  &   739749  &  12.01  &  10.67  &    -  \\
g1.12e12  &  1977112  &   564785  &  1222359  &  12.05  &  10.90  &    -  \\
g1.92e12  &  4018048  &  1200667  &  2467821  &  12.37  &  11.20  &    -  \\
g2.79e12  &  5598386  &  1800095  &  3099962  &  12.55  &  11.29  &    -  \\
\hline
g1.26e12  &  1253780  &   563159  &   462669  &  12.03  &  10.46  &   7.55  \\
g1.27e12  &  1150753  &   471966  &   605247  &  11.94  &  10.57  &   7.84  \\
g1.55e12  &  1469159  &   600775  &   679269  &  12.06  &  10.61  &   7.86  \\
g1.62e12  &  1063633  &   533023  &   327877  &  12.00  &  10.31  &   7.61  \\
g2.37e12  &  2159202  &  1098881  &   881920  &  12.30  &  10.73  &   8.30  \\
g2.71e12  &  2164500  &  1195318  &   596766  &  12.35  &  10.55  &   8.19  \\
g3.74e12  &  3397933  &  1715201  &  1182008  &  12.51  &  10.83  &   8.47  \\
\hline
g4.41e12  &   463537  &   250487  &   204647  &  12.55  &  10.83  &   8.84  \\
g4.55e12  &   503192  &   270807  &   164411  &  12.60  &  10.88  &   8.69  \\
g4.81e12  &   594745  &   306880  &   211336  &  12.66  &  11.01  &   8.70  \\
g4.84e12  &   523621  &   267346  &   185147  &  12.60  &  10.93  &   8.66  \\
g5.22e12  &   592556  &   313138  &   187130  &  12.67  &  10.96  &   8.58  \\
g5.41e12  &   644421  &   290846  &   352249  &  12.62  &  11.23  &   9.14  \\
g5.53e12  &   547812  &   306764  &   183274  &  12.65  &  10.83  &   8.69  \\
g6.53e12  &   591850  &   276231  &   266355  &  12.61  &  11.10  &   8.73  \\
g6.57e12  &   763439  &   356974  &   303556  &  12.73  &  11.15  &   8.75  \\
g6.86e12  &   838063  &   430471  &   303160  &  12.80  &  11.03  &   8.90  \\
g6.70e12  &  1290678  &   431794  &   693967  &  12.83  &  11.54  &   8.83  \\
g7.50e12  &   648889  &   387253  &   259728  &  12.74  &  11.08  &   9.32  \\
g7.55e12  &   860302  &   433196  &   262410  &  12.81  &  10.95  &   8.79  \\
g7.71e12  &   823620  &   415601  &   312032  &  12.79  &  11.17  &   8.94  \\
g7.92e12  &  1323285  &   446337  &   731566  &  12.84  &  11.55  &   9.06  \\
g8.08e12  &  1003594  &   486702  &   364796  &  12.86  &  11.22  &   9.00  \\
g8.45e12  &   779820  &   368069  &   308291  &  12.74  &  11.17  &   8.80  \\
g1.05e13  &  1146928  &   573974  &   489758  &  12.92  &  11.34  &   9.18  \\
g1.14e13  &  1354041  &   646068  &   532474  &  12.98  &  11.41  &   9.16  \\
g1.17e13  &  2085075  &   889400  &   782176  &  13.14  &  11.32  &   8.92  \\
g1.25e13  &  1217182  &   673068  &   367429  &  13.00  &  11.18  &   9.18  \\
g1.33e13  &  1918776  &   809672  &   751999  &  13.09  &  11.58  &   9.24  \\
g1.44e13  &  2836084  &   876310  &  1474963  &  13.14  &  11.86  &   8.94  \\
g1.54e13  &  1456656  &   722348  &   500173  &  13.03  &  11.36  &   9.17  \\
g1.57e13  &   889955  &   436443  &   310259  &  12.81  &  11.16  &   8.85  \\
g1.63e13  &  2089314  &   953300  &   684168  &  13.16  &  11.43  &   9.25  \\
g1.87e13  &  2278215  &  1101553  &   636663  &  13.23  &  11.39  &   9.29  \\
g2.02e13  &  1947099  &  1056092  &   840843  &  13.18  &  11.59  &   9.59  \\
g2.07e13  &  3177787  &  1497578  &  1020594  &  13.36  &  11.47  &   9.45  \\
g2.10e13  &  2972176  &  1249734  &  1072224  &  13.29  &  11.68  &   9.33  \\
g2.11e13  &  2434943  &  1250655  &   955862  &  13.26  &  11.63  &   9.45  \\
g2.20e13  &  1437878  &   847511  &   578127  &  13.08  &  11.41  &   9.46  \\
g2.37e13  &  2872994  &  1497621  &  1030754  &  13.34  &  11.59  &   9.54  \\
g2.58e13  &  3951084  &  1782390  &  1347000  &  13.44  &  11.66  &   9.47  \\
g3.26e13  &  4333704  &  2100752  &  1118595  &  13.51  &  11.45  &   9.52  \\
g3.42e12  &  2620934  &  1068134  &  1313357  &  12.30  &  10.91  &   8.12  \\
g3.78e13  &  3462973  &  1607283  &  1074439  &  13.39  &  11.51  &   9.45  \\
g3.89e13  &  5426667  &  2281160  &  1751607  &  13.55  &  11.81  &   9.70  \\
\hline
g5.41e12  &  1265524  &   337582  &   778618  &  12.73  &  11.59  &    -  \\
g6.86e12  &  1656812  &   453120  &   975692  &  12.86  &  11.69  &    -  \\
g7.50e12  &  1478432  &   443431  &   834490  &  12.85  &  11.62  &    -  \\
g7.92e12  &  1546499  &   455160  &   925496  &  12.85  &  11.65  &    -  \\
g1.05e13  &  2286298  &   658752  &  1301680  &  13.02  &  11.75  &    -  \\
g1.44e13  &  2738604  &   878724  &  1374316  &  13.14  &  11.83  &    -  \\
  \end{tabular}
\end{table*}

\bsp
\label{lastpage}
\end{document}